\documentclass[pre,twocolumn]{revtex4-1}

\usepackage{amssymb,amsmath}
\usepackage{epsfig}
\usepackage{bm}
\usepackage{color}
\usepackage{graphicx}% Include figure files

\definecolor{dark-green}{rgb}{0.2,0.6,0.3}

\newcommand{\vicente}[1]{{ #1}}
\newcommand{\fran}[1]{{ #1}}

\newcommand\beq{\begin{equation}}
\newcommand\eeq{\end{equation}}
\newcommand\beqa{\begin{eqnarray}}
\newcommand\eeqa{\end{eqnarray}}
\newcommand{\dd}{\text{d}}

\begin{document}

\title{Hydrodynamic granular segregation induced by boundary heating and shear}
\author{Francisco Vega Reyes}
\email{fvega@unex.es} \homepage{http://www.unex.es/eweb/fisteor/fran/}
\affiliation{Departamento de F\'isica, Universidad de Extremadura, 06071 Badajoz, Spain}
\author{Vicente Garz\'o}
\email{vicenteg@unex.es}\homepage{http://www.unex.es/eweb/fisteor/vicente/}
\affiliation{Departamento de F\'isica, Universidad de Extremadura, 06071 Badajoz, Spain}
\author{Nagi Khalil}
\email{nagi@us.es}
\affiliation{Departamento de F\'isica, Universidad de Extremadura, 06071 Badajoz, Spain}

% \pacs{05.20.Dd, 45.70.Mg, 51.10.+y, 05.60.-k}

\begin{abstract}

Segregation induced by a thermal gradient of an impurity in a driven low-density granular gas is studied. The system is enclosed between two parallel walls from which we input thermal energy to the gas. We study here steady states occurring when the inelastic cooling is exactly balanced by some external energy input (stochastic force or viscous heating), resulting in a uniform heat flux. A segregation criterion based on Navier-Stokes granular hydrodynamics is written in terms of the tracer diffusion transport coefficients, whose dependence on the parameters of the system (masses, sizes and coefficients of restitution) is explicitly determined from a solution of the inelastic Boltzmann equation. The theoretical predictions are validated by means of Monte Carlo and molecular dynamics simulations, showing that Navier-Stokes hydrodynamics produces accurate segregation criteria even under strong shearing and/or inelasticity.
\end{abstract}

\date{\today}
\maketitle

\section{Introduction}
\label{sec1}

Segregation of particulate matter is a problem with important practical applications \cite{JNB96} \fran{that} are of obvious interest for industry, technology and bio-medical \fran{sectors \cite{RSPS87,PFBBB91,JNB96,S04}. In fact, not only granular matter is one of the most used materials in human applications \cite{dG99} but also granular dynamics is present in many biophysics problems \cite{SCCHD12} and in a variety of ecosystems \cite{B54,H13}. In particular, in a number of applications (fluidized beds, just to put one example \cite{DHC77}) is of interest to separate grains according to their mass and/or size, the so-called \textit{segregation} process, or also the opposite effect (\textit{mixing} of disparate particles \cite{PFBBB91}).} Either result may be needed for industry applications. It is thus of interest in a multidisciplinary context to develop theoretical criteria  capable to predict the behavior of granular segregation. \fran{Quite surprisingly, and in spite of the high economic impact that efficient grain segregation has on many industries \cite{dG99} and also in spite of extensive observation of grain segregation phenomena, no comprehensive transport theory has been extensive and systematically tested against measurements in computer or laboratory experiments \cite{K04,NJP}. We perform this task in the present work, presenting strong evidence of validation of one of the scarce complete kinetic theories on granular gas segregation.}

The use of hydrodynamic transport theories has been extended to granular gases \cite{D01,G03}; i.e.
sparse granular systems \fran{where the dynamics is dominated by particle collisions \cite{H86}.} \fran{Several} works have studied
thermal diffusion segregation from kinetic theories for dilute \cite{BRM05,G06,BKD11}
and \fran{moderately} dense \cite{JY02,TAH03,G08,G11} granular gases. \fran{In this paper}, we show that Navier-Stokes (NS) hydrodynamics derived from kinetic theory accurately predicts segregation for steady granular flows, even under strong collisional dissipation. The agreement covers an entire flow class characterized by having a uniform heat flux. Moreover, since this class of flows has elements with and without shear \cite{VSG10}, we show that our NS segregation criterion works well for both thermal-induced and shear-induced granular gas segregation. It must be remarked that our segregation criterion involves the set of diffusion transport coefficients of the impurity mass flux. The dependence of these transport coefficients on the parameter space of the problem (masses, sizes and coefficients of restitution) is explicitly determined by solving the inelastic Boltzmann equations for the system (impurity plus granular gas) by means of the Chapman-Enskog method \cite{CC70} adapted to dissipative dynamics. Theoretical results are compared with numerical solutions of the kinetic equations of the system (direct simulation Monte Carlo method, DSMC)  and also with molecular dynamics (MD) simulations. Good agreement is found between the three independent solutions. To the best of our knowledge, the comparison carried out in this paper can be considered as one of the most stringent quantitative assessments of kinetic theory to date for conditions of practical interest for thermal diffusion segregation in granular gases.

\vicente{The plan of the paper is as follows. In Sec.\ \ref{sec2} we offer a \fran{brief description} of the set of
inelastic Boltzmann kinetic equations for the granular gas and the impurities. \fran{Section \ref{sec3} presents
a description of the steady flows over which we analyze granular impurity segregation (driven states with uniform heat flow).
There is also a discussion on the derivation of the thermal diffusion factor $\Lambda$, a magnitude that provides the
segregation criterion. By using a \fran{NS} hydrodynamic description, $\Lambda$ is expressed in terms of the impurity diffusion coefficients, whose explicit dependence on the masses, sizes and coefficients of restitution is obtained from a Chapman-Enskog solution of the inelastic Boltzmann and Boltzmann-Lorentz kinetic equations}. \fran{The reliability of the NS thermal diffusion factor is assessed against computer simulations in Sec.\ \ref{sec4}. A remarkable agreement between kinetic theory and simulations is found for conditions of practical interest (strong dissipation and particle dissimilarity). Two different types of simulations are performed for this task: a numerical solution of the Boltzmann equations from the DSMC method and MD simulations.} In Sec.\ \ref{sec5} \fran{we propose a hypothetical} granular segregation laboratory experiment. With this we expect to help future experimental research and applications to use our theoretical results for segregation. The paper is closed in Sec.\ \ref{sec6} with a brief discussion of the results.}

%\vicente{Section II is COMPLETELY NEW}

\section{Kinetic theory description}
\label{sec2}

%\subsection{Boltzmann and Boltzmann-Lorentz kinetic equations}

We consider a set of identical inelastic smooth hard  disks ($d=2$) or spheres ($d=3$) of mass $m$ and diameter $\sigma$  at low density (\textit{granular gas}). Particles collide loosing a fraction of their kinetic energy after collisions. The degree of inelasticity is characterized by the (constant) coefficient of normal restitution $\alpha$, which ranges from 1 (elastic collision, no energy loss) to 0 (perfectly inelastic collision) \cite{D01}. In our system there is also another set of particles with mass $m_0$ and diameter $\sigma_0$, in general different from the values $m$ and $\sigma$, respectively. The relative concentration of this second set of particles is very small compared to that of the other (granular gas) component  and for this reason we call it impurity or intruder. Collisions between impurity-gas particles are also inelastic and characterized by a coefficient of restitution $\alpha_0$.

Since the relative concentration of impurity particles is very small compared to that of the other (solvent or excess) component, one can assume that the state of the granular gas is not affected by the presence of impurity. Moreover, the collisions among impurity particles themselves can be neglected as compared with their interactions with the particles of the granular gas. At a kinetic theory level, this implies that the velocity distribution function $f(\mathbf{r}, \mathbf{v},t)$ of the gas particles obeys the closed (inelastic) Boltzmann equation while the velocity distribution function $f_0(\mathbf{r}, \mathbf{v},t)$ of the impurity particles obeys the (linear) Boltzmann-Lorentz equation.

The Boltzmann kinetic equation for the granular gas is given by \cite{BP04}
\beq
\label{2.1}
\partial_t f+\mathbf{v}\cdot \nabla f+{\cal F} f=J[\mathbf{v}|f,f],
\eeq
where
\beqa
\label{2.2}
J\left[\mathbf{v}_1|f, f\right] &=&\sigma^{d-1}\int
\dd\mathbf{v}_{2}\int \dd\widehat{\boldsymbol {\sigma}}\Theta
(\widehat{\boldsymbol {\sigma}}\cdot \mathbf{g})(\widehat{
\boldsymbol {\sigma }}\cdot \mathbf{g})\nonumber\\
& & \times\left[ \alpha^{-2}f(\mathbf{v}_{1}^{\prime
})f(\mathbf{v}_{2}^{\prime})-f(\mathbf{v}_{1})f(\mathbf{v}_{2})\right]
\eeqa
is the (inelastic) Boltzmann collision operator. In Eq.\ \eqref{2.1}, $\cal{F}$ is an operator representing the effect of an external force that injects energy into the granular gas allowing it to reach a steady state. Moreover, $\widehat{\boldsymbol {\sigma}}$ is a unit vector along the line joining the centers of the two colliding spheres, $\Theta$ is the Heaviside step function, ${\bf g}={\bf v}_{1}-{\bf v}_{2}$ is the relative velocity, and the primes on the velocities denote the initial values $\{\mathbf{v}_1', \mathbf{v}_2'\}$ that lead to $\{\mathbf{v}_1, \mathbf{v}_2\}$ following a binary collision:
\begin{equation}
\label{2.3}
\mathbf{v}_{1,2}'=\mathbf{v}_{1,2}\mp\frac{1}{2}(1+\alpha^{-1})(\widehat{{\boldsymbol {\sigma }}}\cdot
\mathbf{g})\widehat{\boldsymbol {\sigma}}.
\end{equation}

At a hydrodynamic level, the relevant quantities are the density $n$, the flow velocity \fran{$\mathbf{u}$}, and the granular temperature $T$. They are defined as
% creo que es mejor usar en la medida de lo posible siempre la misma notacion,
\begin{equation}
\label{2.4}
n=\int\;\dd\mathbf{v}f(\mathbf{v}),
\end{equation}
\beq
\label{2.5}
\fran{\mathbf{u}=\frac{1}{n}\int\;\dd\mathbf{v}\mathbf{v} f(\mathbf{v}),}
\eeq
\begin{equation}
\label{2.6}
T=\frac{m}{d n}\; \int\; \dd \mathbf{v}\; V^2 f(\mathbf{v}),
\end{equation}
where \fran{$\mathbf{V}=\mathbf{v}-\mathbf{u}$} is the peculiar velocity. The macroscopic balance equations for number density $n$, momentum density \fran{$m{\bf u}$}, and energy density $\frac{d}{2}nT$ follow directly from Eq.\ ({\ref{2.1}) by
multiplying with $1$, $m{\bf V}$, and $\frac{1}{2}m V^2$ and integrating over ${\bf v}$:
\begin{equation}
\label{2.7} \fran{D_{t}n+n\nabla \cdot {\bf u}=0\;,}
\end{equation}
\begin{equation}
\label{2.8} \fran{D_{t}{\bf u}+(mn)^{-1}\nabla \cdot {\sf P}={\bf 0}\;,}
\end{equation}
\begin{equation}
\label{2.9} \fran{D_{t}T+\frac{2}{dn}\left(\nabla \cdot {\bf q}+P_{ij}\nabla_j u_{i}\right) =-\left(\zeta-\sigma_T\right) T\;.}
\end{equation}
Here, $D_{t}=\partial _{t}+{\bf u}\cdot \nabla$ is the material time derivative,
\beq
\label{2.10}
P_{ij}=m \; \int\; \dd \mathbf{v}\; V_i V_j f(\mathbf{v}),
\eeq
is the pressure tensor,
\beq
\label{2.11}
\mathbf{q}=\frac{m}{2} \; \int\; \dd \mathbf{v}\; V^2 \mathbf{V} f(\mathbf{v}),
\eeq
is the heat flux, and
\beq
\label{2.12}
\zeta=-\frac{m}{d n T} \; \int\; \dd \mathbf{v}\; V^2 J[\mathbf{v}|f,f]
\eeq
is the cooling rate characterizing the rate of energy dissipated in collisions.
In addition,
\beq
\label{2.13}
\sigma_T=-\frac{m}{d n T} \; \int\; \dd \mathbf{v}\; V^2 {\cal F}f(\mathbf{v})
\eeq
is the source term measuring the rate of heating due to the external force. In Eqs.\ \eqref{2.7}--\eqref{2.9}, it is assumed that the external driving does not change the number of particles or the momentum, i.e.,
\beq
\label{2.14}
\int\; \dd \mathbf{v}\; {\cal F}f(\mathbf{v})=\int\; \dd \mathbf{v}\; V_i {\cal F}f(\mathbf{v})=0.
\eeq

The Boltzmann-Lorentz kinetic equation for impurities is given by \cite{BKD11}
\beq
\label{2.15}
\partial_t f_0+\mathbf{v}\cdot \nabla f_0+{\cal F} f_0=J[\mathbf{v}|f_0,f],
\eeq
where the collision operator $J_0[\mathbf{v}|f_0,f]$ is
\beqa
\label{2.16}
J_0\left[\mathbf{v}_1|f, f\right] &=&\overline{\sigma}^{d-1}\int
\dd\mathbf{v}_{2}\int \dd\widehat{\boldsymbol {\sigma}}\Theta
(\widehat{\boldsymbol {\sigma}}\cdot \mathbf{g})(\widehat{
\boldsymbol {\sigma }}\cdot \mathbf{g})\nonumber\\
& & \times\left[ \alpha_0^{-2}f_0(\mathbf{v}_{1}^{\prime
})f(\mathbf{v}_{2}^{\prime})-f_0(\mathbf{v}_{1})f(\mathbf{v}_{2})\right].\nonumber\\
\eeqa
Here, $\overline{\sigma}=(\sigma_0+\sigma)/2$ and the precollisional velocities are given by
\begin{equation}
\label{2.17}
\mathbf{v}_{1}'=\mathbf{v}_{1}-\frac{m}{m_0+m}(1+\alpha_0^{-1})(\widehat{{\boldsymbol {\sigma }}}\cdot
\mathbf{g})\widehat{\boldsymbol {\sigma}},
\end{equation}
\begin{equation}
\label{2.18}
\mathbf{v}_{2}'=\mathbf{v}_{2}+\frac{m_0}{m_0+m}(1+\alpha_0^{-1})(\widehat{{\boldsymbol {\sigma }}}\cdot
\mathbf{g})\widehat{\boldsymbol {\sigma}}.
\end{equation}}

\begin{figure}
\begin{center}
\includegraphics[width=1.0 \columnwidth,angle=0]{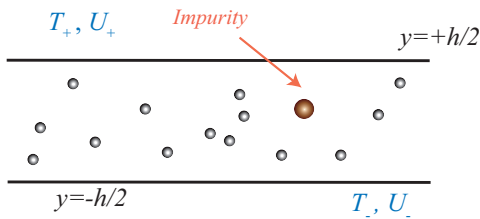}
\caption{\fran{We observe here a sketch of the system. In this case the impurity is represented as a particle bigger than the gas particles. Energy is input in the system from the boundaries, which consist of two infinite parallel walls with temperature and shear sources.}} \label{fig0}
\end{center}
\end{figure}

Impurities may freely lose or gain momentum and energy when they collide with particles of the gas and, hence the number density
\begin{equation}
\label{2.19}
n_0=\int\;\dd\mathbf{v}f_0(\mathbf{v})
\end{equation}
is the \emph{only} collisional invariant of $J_0[f_0,f]$. Its conservation equation is
\beq
\label{2.20}
D_t n_0+n_0 \nabla \cdot \mathbf{u}=-\frac{\nabla \cdot {\mathbf j}_0}{m_0},
\eeq
where
\beq
\label{2.21}
\mathbf{j}_0=m_0 \; \int\; \dd \mathbf{v}\; \mathbf{V} f_0(\mathbf{v})
\eeq
is the mass flux of impurities relative to the (local) \fran{flow $\mathbf{u}$}. Upon deriving Eq.\ \eqref{2.20}, we have assumed that the external driving force does not change the number of impurities.

Apart from the fluxes, a relevant quantity at a kinetic level is the temperature of the impurity (or tracer temperature) $T_0$. It is defined as
\beq
\label{2.22}
T_0=\frac{m_0}{d n_0}\; \int\; \dd \mathbf{v}\; V^2 f_0(\mathbf{v}).
\eeq
The partial temperature $T_0$ measures the mean kinetic energy of the impurity. This quantity is in general different from the granular temperature $T$.

\section{Thermal diffusion segregation in driven steady states}
\label{sec3}

\fran{\subsection{Hydrodynamic profiles}}

\fran{The main goal of this paper is to study the conditions for which, in driven steady states, impurity particles }tend to separate (segregate) from the granular gas. Let us first briefly describe the physical situation we are interested in. The system is enclosed between two infinite parallel walls at $y=-h/2$ and $y=+h/2$ from which is heated and (optionally) also sheared. The upper wall input temperatures and velocities are\fran{, respectively,} $T_+$ and $U_+$ while the lower wall inputs are $T_-$ and \fran{$U_-$}. Here, we will consider always $T_+>T_-$. Also, we study situations where the \fran{energy input} from the walls is sufficiently strong at all times so that the influence of gravity is not important. Additionally there is an energy input in the granular gas volume that in our case we model as a white noise \cite{PLMPV98}. A sketch of the geometry of the problem is given in Fig.\ \ref{fig0}.

\fran{In our problem, and for \emph{steady} base states, spatial gradients} occur only along the $y$-direction \cite{VSG10}. More specifically, we consider in this work only states with \emph{uniform} \fran{heat flux}. In this case, the energy balance equation \eqref{2.9} reads \cite{VSG13}
\begin{equation}
\label{qbal}
\frac{\partial q_y}{\partial y}=-\frac{d}{2}nT(\zeta-\sigma_T)-P_{xy} \frac{\partial U_x}{\partial y} =0.
\end{equation}
\vicente{Here, viscous heating \fran{$-P_{xy}\partial_y u_x$ and volume energy input $(d/2)nT\sigma_T$ balance the inelastic cooling term $-(d/2)nT\zeta$. We will consider in this work two types of uniform heat flux steady flows.}}

(a) \textit{\underline{Case \text{I}}:} $U_\pm=0$ (no shear), which implies \fran{$u_x(y)=0$} in Eq.\ \eqref{qbal}.
\fran{As we said, we assume here} that the system is driven by means of a stochastic Langevin force representing a Gaussian white noise \cite{WM96}. The covariance of the stochastic acceleration $\xi^2$ \cite{PLMPV98} is chosen to be the same for both species (impurity and gas particles) \cite{GV12}. In the context of the kinetic equations \eqref{2.1} and \eqref{2.15}, this external force is represented by a Fokker-Planck operator \cite{NE98} of the form ${\cal F}\equiv -\frac{1}{2}\xi^2\partial^2/\partial v^2$. In this case, the production of energy term is  $\sigma_T=m\xi^2/T$. The stochastic external forcing is frequently used in computer simulations \cite{PLMPV98,CLH00,KSSAOB05,CBMNS06,FAZ09,KSZ10} and has been also proved experimentally \cite{KSSAOB05,OLDLD04}.

(b) \textit{\underline{Case \text{II}}:}  No \fran{volume} driving \fran{ ($\sigma_T=0$) and boundary shear ($U_-U_+\neq0$)}, i.e., when both walls are in relative motion (sheared granular gas). In this case, inelastic cooling is compensated by viscous heating.

\fran{Note that in both Cases I and II inelastic cooling may be achieved locally for all points in the system \cite{GM02, VSG10}.
It is also important to remark that the bulk hydrodynamic profiles of the above situations are not simple since for the uniform heat flux flow class the hydrodynamic profiles fulfill \cite{VSG10}
\begin{equation}
T(y)\propto y^{2/3},\quad u_x(y)\propto y^{2/3},
\label{profiles}
 \end{equation}
the latter applying only in the sheared system. Furthermore, as} we noted in a previous work \cite{VSG10}, with  uniform heat flux and at fixed $T_\pm$ at the walls, the dimensionless temperature profile (scaled with proper units) is \emph{universal}, namely, \fran{independent} of the shearing or inelasticity conditions. As a consequence,  Cases I and II share the same $T(y)$-profiles. This surprising result is a consequence of the applicability of hydrodynamics to granular gases, even for strong dissipation. \fran{Moreover, it allows us for obtaining segregation conditions in applications in a much more simple way, as we see below.}

\fran{\subsection{Segregation criterion}}

As said before, the only relevant space direction in our problem is orthogonal to the walls ($y$ axis). In this case, the thermal diffusion factor $\Lambda$ characterizes the amount of segregation parallel to the thermal gradient. It is defined through the relation \cite{GI52}
\begin{equation}
\label{Lambda}
-\Lambda\frac{\partial\ln T}{\partial y}=\frac{\partial \ln (n_0/n)}{\partial y}.
\end{equation}
\fran{It is interesting to note that uniform heat flux profiles, from Eqs.\ \eqref{profiles}, do not show an absolute minimum or maximum in the bulk of the fluid. Thus, no change of segregation behavior, as given by Eq.\ \eqref{Lambda}, occurs in the system as a function of space coordinates. This is important for applications since the uniform heat flux flow class yields a clear and unique segregation behavior for a given experimental configuration.}

When the impurity is larger than the gas particles ($\sigma_0>\sigma$) and concentrates in opposite direction to gravity, the system beholds the so-called Brazil nut effect \cite{RSPS87}. Otherwise, the so-called reverse Brazil nut effect is observed \cite{S04}. In this work we consider arbitrary values for the mass $\mu\equiv m_0/m$ and size $\omega \equiv \sigma_0/\sigma$ ratios and besides gravity effects are negligible. Thus, we will not use this terminology and will just refer to the sign of $\Lambda$. For our boundary conditions and taking into account Eq.\ \eqref{Lambda}, when $\Lambda<0$ the impurities drift to the hot wall (upwards) while if $\Lambda>0$ the impurities go down to the cold wall (downwards). Segregation changes sign (the region of preference of the impurity switches) at the marginal points $\Lambda=0$. Exactly at $\Lambda=0$ points, the impurity has no region of its preference and mixing occurs. Moreover, as simulations clearly show, the factor $\Lambda$ is \emph{uniform} in the bulk region. Consequently, the segregation criterion derived from the condition $\Lambda=0$ is a global feature of the bulk domain and is not restricted to specific regions of the system.

\vicente{In the steady state, the momentum balance equation \eqref{2.8} implies $P_{xy}=\text{const.}$ and $P_{yy}=\text{const.}$ In addition, since $\nabla \cdot \mathbf{u}=0$ in Cases I and II, then the mass flux $j_{0,y}$ vanishes in the steady state according to Eq.\ \eqref{2.20}. To close the problem of determining $\Lambda$ one needs a constitutive equation for the mass flux $j_{0,y}$. In the first order in the spatial gradients (NS approximation), $j_{0,y}$ is given by \cite{G08}
\begin{equation}
\label{massflux}
j_{0,y}=-\frac{m_0^2}{\rho}D_0\frac{\partial n_0}{\partial y}-\frac{m_0m}{\rho}D\frac{\partial n}{\partial y}-\frac{\rho}{T}D^T\frac{\partial T}{\partial y},
\end{equation}
where $D_0$, $D$, and $D^T$ are the impurity diffusion transport coefficients. The condition $j_{0,y}=0$ along with Eq.\ \eqref{massflux} yields
\begin{equation}
\label{Lrel}
\Lambda=\frac{D^{T*}-D_0^*-D^*}{D_0^*},
\end{equation}
where we have introduced the reduced transport coefficients $D^{T*}=(\rho\nu/n_0T)D^T$,
$D_0^*=(m_0^2\nu/\rho T)D_0$, and $D^*=(m_0\nu/n_0T)D$. Here, $\nu=n\sigma^{d-1}\sqrt{2T/m}$ is an effective collision frequency.

As for elastic collisions, the (reduced) diffusion coefficients $D^{T*}$, $D_0^*$, and $D^*$ are given in terms of the solutions of a set of coupled linear integral equations. The standard method consists of approximating the unknowns by Maxwellians (at different temperatures) times truncated Sonine polynomial expansions. In the most simple approximation, only the lowest Sonine polynomial (first Sonine approximation) is retained and the result when the gas is heated by the stochastic force (Case I) is \cite{G08}
\beq
\label{3.2}
D_0^*[1]=\frac{\chi}{\nu_D^*}, \quad  D^{T*}[1]=\frac{\chi-\mu}{\nu_D^*}, \quad D^*[1]=-\frac{\mu}{\nu_D^*},
\eeq
where $D_0^*[1]$, $D^{T*}[1]$ and $D^*[1]$ refer to the first Sonine approximation to $D_0^*$, $D^{T*}$, and $D^*$, respectively. In addition, $\chi=T_0/T$ is the temperature ratio and $\nu_D^*$ is a known collision frequency. Substitution of the expressions \eqref{3.2} into Eq.\ \eqref{Lrel} yields $\Lambda=0$, namely, the first Sonine approximation does not predict segregation ($\Lambda=0$) in the driven states analyzed here for dilute systems. Note that the first Sonine solution to $\Lambda$ yields segregation ($\Lambda \neq 0$) for dense systems \cite{GV12}. Thus, as for binary elastic mixtures \cite{KCL87}, one has to determine the diffusion coefficients by considering the second Sonine approximation (two polynomials in the Sonine polynomial expansion) to the distribution functions. The explicit second Sonine forms $D_0^*[2]$, $D^{T*}[2]$ and $D^*[2]$ for a dilute system are given in the Appendix. }

Since we are considering a unique $T(y)$ profile (once $T_\pm$ are fixed at the boundaries), then  (see Eq.\  \eqref{Lambda}) a unique segregation profile $n_0(y)/n(y)$ also results, except for the constant factor $\Lambda$. Furthermore, $\Lambda$  is usually employed to describe granular segregation problems but in fact it is only when $T(y)$ uniqueness applies (as in here) when $\Lambda$ is purposeful. Thus, we only need to provide an accurate value of the $\Lambda\equiv \Lambda(\Xi,\chi)$ function, in order to properly describe the segregation behavior. Here, $\Xi\equiv \left\{\alpha,\alpha_0,\mu,\omega\right\}$ denotes the set of mechanical properties of the system. It must be noted that the temperature ratio $\chi$ is also uniform for steady base states in our geometry \cite{GV10}.

\begin{figure}
\includegraphics[height=5.0cm]{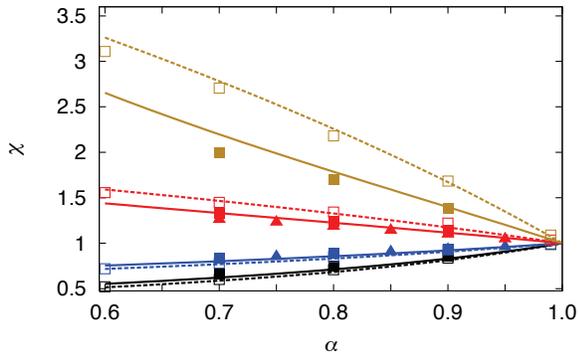}
\caption{Temperature ratio vs. the (common) coefficient of restitution $\alpha=\alpha_0$ for $\omega=1$ and several values of relative mass $\mu$:  $\mu=1/4$ (black), $\mu=1/2$ (blue), $\mu=2$ (red), and $\mu=4$ (brown). The dashed and solid lines refer to the theoretical values obtained for Case I (no shear) \cite{GV12} and Case II (shear) \cite{G02}, respectively. Symbols stand for simulations: open symbols for Case I and solid symbols for Case II (triangles for MD and squares for DSMC). All figures in this work stand for spheres ($d=3$).} \label{fig1}
\end{figure}

We will make the ansatz that the dependence of the $\Lambda$ function (which results from the calculation of the transport coefficients characterizing the mass flux of impurities) on $\Xi$ and $\chi$ is common for all flows in the class, sheared or not, since all of them have common heat flux and temperature profile properties. Thus, for given values of the parameters of the system, $\Lambda$ will be obtained from Eq.\ \eqref{Lrel} by using the shear-independent forms \cite{GV12} of the diffusion coefficients displayed in the Appendix. \vicente{The NS coefficients $D_0^*[2]$, $D^{T*}[2]$ and $D^*[2]$  are functions of the temperature ratio $\chi$. Nevertheless, the value of $\chi$ strongly depends not only on the mechanical properties $\Xi$ but also on shearing conditions \cite{G02}. Therefore, in Case I, the temperature ratio $\chi$ fulfills the condition \cite{BT02, DHGD02} %fran estas referencias no estan en el bib que me has mandado
\begin{equation}
\label{3.3}
\chi_I \zeta_0^*=\mu \zeta^*,
\end{equation}
while in Case II, $\chi$ is given by
\beq
\label{3.4}
\chi_{II}=\frac{\zeta^* P_{0,xy}^*}{\zeta_0^* P_{xy}^*}.
\eeq
In Eqs.\ \eqref{3.3} and \eqref{3.4}, the cooling rates $\zeta^*$ and $\zeta_0^*$ (which measure the rate of change of $T_0$) are
\begin{equation}
\label{3.5}
\zeta^*=\frac{\sqrt{2}\pi ^{(d-1)/2}}{d\Gamma \left( \frac{d}{2}\right)}  (1-\alpha^2),
\end{equation}
\begin{eqnarray}
\label{3.6}
\zeta_0^*&=&\frac{4\pi ^{(d-1)/2}}{d\Gamma \left( \frac{d}{2}\right)}
\left(\frac{1+\omega}{2}\right)^{d-1}\frac{1}{1+\mu}\left( \frac{\mu+\chi}{\mu}
\right)^{1/2}\nonumber\\
&\times& (1+\alpha _{0})\left[1-\frac{\mu+\chi}{2\chi(1+\mu)}(1+\alpha_0)\right].
\end{eqnarray}

Moreover, in Eq.\ \eqref{3.4}, the pressure tensors of gas particles $P_{ij}$ and impurity $P_{0,ij}=\int\; d\mathbf{v} m_0V_i V_j f_0$ have been determined from Grad's moment method for sheared (or non-Newtonian) base states with uniform heat flux \cite{G02}. This ansatz yields two different $\Lambda$ values: $\Lambda(\Xi,\chi_I)$ for Case I and $\Lambda(\Xi,\chi_{II})$ for Case II. The final expressions for the thermal diffusion factor $\Lambda(\Xi,\chi)$ and the temperature ratios  $\chi_I$ and $\chi_{II}$ are rather involved and may be found in previous works \cite{G02,GV12}.}

\begin{figure}
\includegraphics[height=5.0cm]{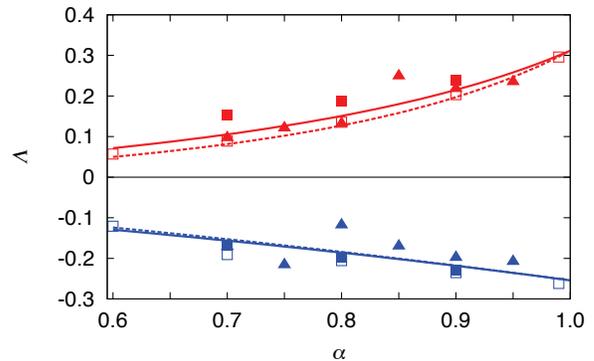}
\caption{Thermal diffusion factor $\Lambda$ vs. the (common) coefficient of restitution $\alpha=\alpha_0$ for $\omega=1$. Two relative mass cases are represented:  $\mu=1/2$ (blue) and $\mu=2$ (red). The dashed and solid lines correspond to the theoretical predictions obtained for Case I and Case II, respectively. Both Cases draw rather similar curves. Also, open and solid symbols stand for simulation data of Case I and Case II (triangles for MD and squares for DSMC), respectively.} \label{fig2}
\end{figure}

\begin{figure}
\includegraphics[height=5.0cm]{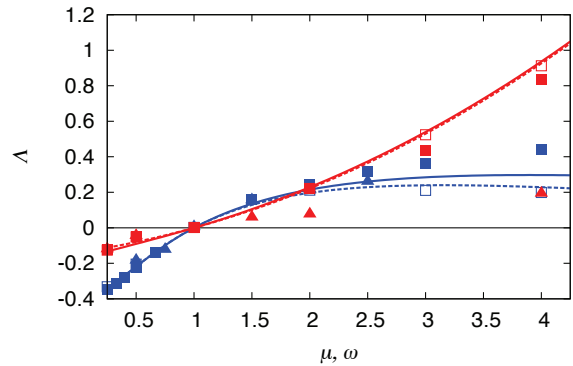}
\caption{Thermal diffusion factor $\Lambda$ vs. relative mass $\mu$ and diameter $\omega$. We represent the case $\alpha=\alpha_0=0.9$: for relative size $\omega=1$ and variable relative mass $\mu$ (blue) and for $\mu=1$ and variable relative size $\omega$ (red). The meaning of the lines and the symbols is the same as that of Fig.\ \ref{fig2}.} \label{fig3}
\end{figure}

\section{Comparison between kinetic theory and computer simulations}
\label{sec4}

In order to canvass our ansatz, we have performed DSMC simulations of the kinetic equations and also MD simulations (event-driven algorithm \cite{L91,PS05}) for hard spheres ($d=3$). With DSMC simulations we validate the NS theory (first order in spatial gradients) compared to the kinetic equations from which it results while with MD simulations we validate the kinetic equation themselves (since MD simulations avoid any assumptions inherent in the kinetic theory, such as molecular chaos hypothesis). All simulations have been performed for $T_+/T_-=5$ and $h=15\overline{\lambda}$, where $\overline{\lambda}=(\sqrt{2}\pi\overline{n}\sigma^2)^{-1}$ is the mean free path and $\overline{n}$ is the average particle density. Thus, there is a unique temperature profile in this work: the one corresponding to conventional Fourier flow (without shearing) for a molecular gas heated from two parallel walls with the same boundary conditions for temperature \cite{VSG10}. The packing fraction in MD simulations is $\phi\simeq 0.0071$, which corresponds to a very dilute gas.

Numerical methods for both types of simulations correspond to the
traditional DSMC for the Boltzmann equation and event-driven algorithms
and they are described in more detail elsewhere \cite{VSG10,PS05,BKD12,CVG13}.
It is worth, however, to comment on the energy inputs in the
simulations. Temperature sources at the walls (Cases I and II) are
modeled as regular hard walls (normal component particle velocity to
the wall is inverted), whereas boundary shear (only for Case II) is
performed by adding the wall velocity to horizontal particle velocity.
Respect to volume forcing (Case I), a random velocity is added to all
particles velocities each simulation step, this random velocity being
drawn from a gaussian distribution function whose typical width is
determined by the white noise intensity $\xi^2$. In this case, this
intensity varies with space coordinate so as to satisfy uniform heat
flux condition \eqref{qbal}.

The dependence of the temperature ratios $\chi_I$ and $\chi_{II}$ on the (common) coefficient of restitution $\alpha=\alpha_0$ is plotted in Fig.\ \ref{fig1} for $\omega=1$ and four values of the mass ratio. Comparison between theory and simulations shows clearly a very good agreement for both Cases I and II. It is also apparent that the values $\chi_I$ and $\chi_{II}$ are quite similar in the different systems, except for the most disparate case $m_0/m=4$. Needless to say, the excellent agreement found here at the level of $\chi$ is an important prerequisite for an accurate theoretical segregation criterion, as we explained.

\begin{figure}
\includegraphics[height=5.0cm]{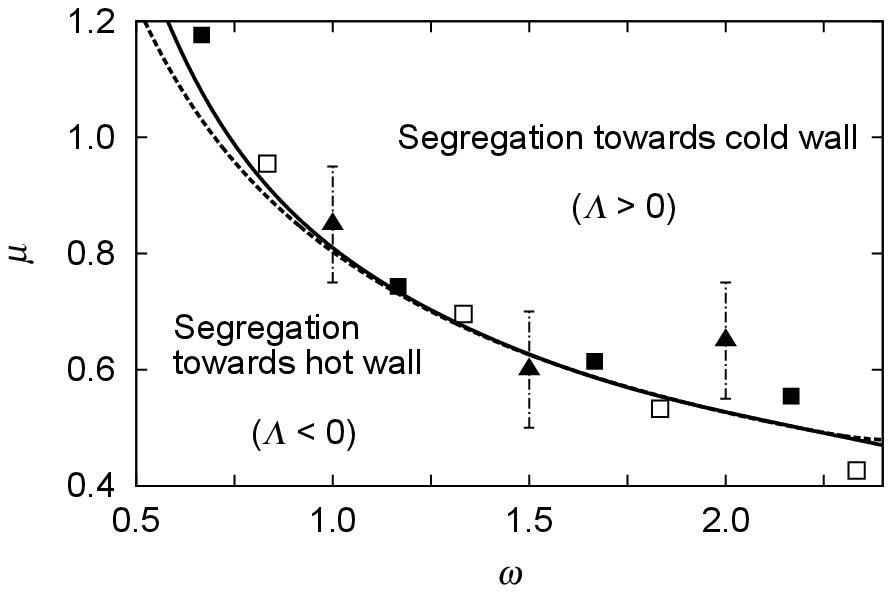}
\caption{Plot of the marginal segregation curve ($\Lambda=0$) for a system with $\alpha=0.9$ and $\alpha_0=0.7$. The dashed and
solid lines stand for the theoretical predictions derived for Cases I and II, respectively.  Open and solid symbols correspond to simulations to Cases I and II (triangles for MD and squares for DSMC), respectively. The error bars have been estimated by using the difference between the values obtained at the bulk region (which are plotted) and those obtained in the whole system.} \label{fig4}
\end{figure}

Next, we compare simulation measurements of $\Lambda$ with the NS theoretical expressions \cite{GV12} for $\Lambda(\Xi,\chi_I)$ and $\Lambda(\Xi,\chi_{II})$. The quantitative agreement here is also very good, as we show in Figs.\ \ref{fig2} and \ref{fig3}, except in the region of large impurities, where the agreement is qualitative. In Fig.\ \ref{fig2}, $\Lambda$ vs. inelasticity is analyzed whereas we study $\Lambda$ vs. the mass and size ratios in Fig.\ \ref{fig3}. We can observe from Figs.\ \ref{fig2} and \ref{fig3} that the dependence of both $\Lambda(\Xi,\chi_I)$ and $\Lambda(\Xi,\chi_{II})$ on $\alpha$ (Fig.\ \ref{fig2}) and $\mu$ or $\omega$ (Fig.\ \ref{fig3}) is actually quite similar. It is also particularly noticeable from Figs.\ \ref{fig2} and \ref{fig3} that kinetic theory reproduces very well the change of sign of $\Lambda$ for Cases I and II.

An accurate prediction of the marginal or critical points $\Lambda=0$ is crucial for applications. For this reason, a phase diagram delineating the regions between $\Lambda>0$ and $\Lambda<0$ in the $(\omega,\mu)$-plane is shown in Fig.\ \ref{fig4} for $\alpha=0.9$ and $\alpha_0=0.7$. We observe first that both Cases draw rather similar theoretical curves in all the range of values of $\mu$ and $\omega$ explored. Moreover, it is quite apparent that the agreement between theory and simulation is excellent, even for a relatively strong degree of inelasticity of impurity-gas particles. Also, the points corresponding to Case II are under strong shearing \cite{VSG10}, and even in that case (which goes beyond the NS description) the agreement with theory is good.

\fran{\section{Experimental setup outline}}
\label{sec5}

In this section, we propose an experimental set-up capable of reproducing in a laboratory the physical situation considered in this paper. Therefore, we need to produce a temperature gradient in the vertical direction (gravity's direction). In this case,
we can think of a parallelepiped system of dimensions \fran{$(h/2)\times(h/2)\times h$, being $h$} the dimension in the vertical direction. Top and bottom walls should be perpendicular to the vertical direction and attached to one (or two) accelerometer(s) capable of reproducing accelerations of up to $25~g$ ($g\sim 10~\mathrm{m/s}^2$ is gravity acceleration). In case of using just one accelerometer, producing different wall-particle contact surfaces in each wall may be enough to generate a vertical temperature gradient. This type of configuration has been devised in a large number of theoretical works, but not yet directly compared with similar experimental configurations. For theoretical works see, additionally to the main text, Ref.\ \cite{GZN97} for a study of steady flows, and Ref.\ \cite{JY02} for a study on granular segregation due to a temperature gradient.

\subsection{Dimensions and density}
\label{dim}

First of all, we need the system to be sparse enough, for instance with a packing fraction $\phi=10^{-2}$ or less. We define $\phi$ as
\begin{equation}
\phi=\frac{\pi}{6}n\sigma^3\simeq \frac{1}{2}n\sigma^3,
\label{nu}
\end{equation}
where $n=N/V$ is the particle density, $N$ being the number of particles and \fran{$V=L^3/4$} the volume of the system.

The other requirement is that the particles should not be not too large nor heavy. Stainless steel balls with diameter $\sigma=10^{-3}$~m is an option. They yield a coefficient of restitution near the quasielastic limit ($\alpha\simeq 0.95$) \cite{L99,G03}. Stainless steel is also convenient for having small sliding effect (small sliding coefficient $\mu\simeq0.099$) and a roughness ($\beta\simeq0.4$) \cite{L99} in the range of nearly Maxwellian behaviour \cite{VSK13}.

Thus, we will pick the reference values $\phi=10^{-2}$ and $\sigma=10^{-3}$~m, and based on them other relevant magnitudes. Our reference particle size for stainless steel implies a particle mass of approximately  $m=0.028304462$~g \cite{L99,VU08}.

An important point is to reproduce the experiment with different degrees of inelasticity in the binary collisions. It may be also needed an impurity made of a different material if we want to perform experiments with $\alpha_0\neq\alpha$. When using different materials for balls in the experiments, and since our model does not include the effects of roughness, it is convenient that the differences in inelasticity are essentially due to the coefficient of normal restitution $\alpha$, and not to the tangential ($\beta$) or sliding ($\mu$) friction coefficients \cite{FLCA94}. To this respect, the use of metallic balls is always convenient, assuring a relatively small sliding effect ($\mu\simeq 0.1$) and little variation in the tangential coefficient of restitution ($\beta\simeq 0.4$ for most metals) \cite{L99}.

With respect to the ball mass, it has an effect on the thermalization of the system (through the collision frequency) but not directly on the dimensions (via the mean free path). The mean free path $\lambda$ is actually of the order \cite{GZN97} $\lambda \sim (n\sigma^2)^{-1}$. Taking this definition into account, we may rewrite the packing fraction \eqref{nu} as
\begin{equation}
\phi=\frac{1}{2}\lambda^{-1}\sigma.
\label{nulambda}
\end{equation}
Another alternative expression for the packing fraction $\phi$ in terms of the number of particles $N$ is
\begin{equation}
\phi=2N\left(\frac{\sigma}{h}\right)^3.
\label{Nlambda}
\end{equation}

From Eq.\ \eqref{nulambda} we obtain
\begin{equation}
\lambda=\frac{\sigma}{2\phi}=B h, 
\label{lambdaAL}
\end{equation}
where in the second equality $B$ is a proportionality constant necessarily small in order to have \fran{$L\gg\lambda$}. We need this condition in order to get an experimental hydrodynamic region (since boundary layers in gases are typically of the order of $\lambda$) \cite{VSG13,G03}. Thus, from Eq.\ \eqref{lambdaAL} one has 
\begin{equation}
h\gtrsim 5\times 10^{-2} B^{-1} \text{m}.
\label{Lbound}
\end{equation}
In addition, by combining conditions \eqref{Nlambda} and \eqref{Lbound}, and for $\sigma=10^{-3}$~m, we get, approximately, that
\begin{equation}
N\gtrsim 6\times10^{2}B^{-3}.\label{Nbound}
\end{equation}

A reasonable value of $B$ is $B=0.1$, which yields, according to Eq.\ \eqref{lambdaAL} a system 10 times larger than the mean free path, sufficient to obtain a wide hydrodynamic central region, even for very far from equilibrium states \cite{VSG13}. Thus, for $\sigma=10^{-3}$~m and $B=0.1$ we obtain from Eqs.\ \eqref{Lbound} and \eqref{Nbound}, respectively, $h\gtrsim 0.5$~m and $N\gtrsim 6\times 10^5$ particles. These seem to be reasonable values \cite{OU98,VU08,GPT13}, but can be decreased as long as the ratio $N/h^3$ is kept, in order to get always $\phi=10^{-2}$. Nevertheless, the relation \eqref{lambdaAL} must be taken into account that, for the same ball size, packing fraction, and consequently, mean free path [see Eq.\ \eqref{nulambda}], a decrease in $h$ leads to an increase in $B$, which is not convenient, since as we said, $B^{-1}$ is a measure of the hydrodynamic region size. Thus, if we stick with the same ball size, we might not have a wide valid margin for parameter values.

\subsection{Energy Input}
\label{energy}

Regarding with the input acceleration $\Gamma$, and as we said, we would need it to be larger than acceleration of gravity. Therefore, for a vibrating wall with amplitude $A$ and angular frequency $\omega$, we would have
\begin{equation}
\Gamma=A\omega^2=\gamma g,
\end{equation}
where $\gamma$ is a factor larger than 1 and $g$ is the acceleration of gravity. Thus, for $A\sim10^{-3}\mathrm{m}$ (an amplitude similar to ball size, we suggest not larger), and a reasonable frequency $f=80$~Hz, which lies in the range of previous and related experimental works \cite{OU98,VU08,GPT13}, we obtain $\gamma=23$, i.e. $\Gamma\sim 23~g$,
which should be large enough for our purpose. In fact, for $h=0.5$~m as we said and assuming approximately constant particle velocities, then, from $v^2\sim A\omega^2 h$, we achieve an average velocity $v\sim 10$~m/s, which gives an approximate idea of the speed of particles during experiment while shaking at these frequencies. Of course if experiments under no gravity conditions are to be performed, there is no need to reach at such high input accelerations, and if the absence of gravity is limited in laboratory time, we also have the advantage that the steady state is in theory rapidly reached, usually in less than 30 collisions per particle \cite{DUSF}.

\section{Discussion}
\label{sec6}

In this paper, we have shown that NS granular hydrodynamics with the diffusion transport coefficients derived from the \emph{inelastic} Boltzmann kinetic theory describes very well thermal diffusion segregation of an impurity in a \emph{low density} granular gas in a steady state with uniform heat flux. The results reported here contain a systematic validation by means of simulations of kinetic theory of granular segregation. The comparison is exhaustive since it covers a wide range of conditions, including moderate and strong dissipation and particle dissimilarity. Furthermore, we show evidence of \emph{quantitative} agreement between granular hydrodynamics and simulations, for both thermal-induced and shear-induced granular segregation. The focus on uniform heat flux case actually does cover a variety of common situations. For instance, in Case I, the noise intensity may simulate a surrounding molecular gas that fluidizes the granular gas, analogously to a previous work \cite{OLDLD04}. In our case this molecular gas shows the typical Fourier flow temperature profile (since it has uniform heat flux), which seems the most natural situation for a molecular gas heated from two parallel walls. Furthermore, Cases I and II may be relevant also for granular active matter systems, where spontaneous steady flow may appear \cite{SCDHD13}. Our study obviously also applies for pipe granular flows applications, where knowledge of shear-induced segregation is important. \fran{We have proposed a test experiment, that we think is relatively easy to set-up, in order to check our theory.}

\fran{We think it is} of particular interest the fact that segregation has almost the same behavior for Cases I and II (see Figs.\ \ref{fig2}-\ref{fig4}), which in principle could look like very different from each other: in Case I the heat flux balance is produced by a volume stochastic force and the system is not sheared whereas in Case II there are no volume forces and heat flux balance. Of course, since the reduced $T(y)$ profiles are the same for both Cases (for all uniform heat flux flows, as we know \cite{VSG10}), the surprising fact is reduced to the coincidence in the balance of the different diffusion coefficients, as we can see from Eq.\  \eqref{massflux}. Therefore, it seems that the common transport properties that connect all flows with uniform heat flux for a monodisperse granular gas \cite{VSG10} may be extended to impurity segregation behavior. In addition, although our present description is restricted to dilute granular gases, we expect that the main results reported here for thermal diffusion segregation can be extended to higher densities and different flow classes.

\vicente{Given that segregation is one of the most important open challenges in granular flows research, \fran{and granular transport related industries, }the results displayed in the present paper could be of great value not only for experts in kinetic theory for granular gases but also for more applied scientists \fran{and engineers}. In fact, according to the relatively small experimental setup to measure segregation in a granular gas proposed in Sec.\ \ref{sec6}, we think that the reliability of our results could be assessed via a comparison with experimental data since they have a direct application in laboratory experiments.}

\acknowledgments

We acknowledge support of the Spanish Government through grant Nos. FIS2010-16587 (V. G., F. V. R. and N.K.) and MAT2009-12351-C02-02 (F. V. R.). The first grant is partially financed by FEDER funds and by Junta de Extremadura (Spain) through Grant No.\ GRU10158.

\appendix
\section{Explicit expressions for $D_0^*[2]$, $D^{T*}[2]$ and $D^*[2]$}
\label{appA}

The explicit forms of the second Sonine approximations to the diffusion transport coefficients in the low-density limit are displayed in this Appendix. They can be easily obtained from the expressions displayed in Appendix of Ref.\ \cite{GV12} when the  volume fraction vanishes. They are given by
\begin{equation}
\label{a1}
D_{0}^*[2]=\frac{\nu_4^*\chi}{\nu_1^*\nu_4^*-\nu_2^*(\nu_5^*-\zeta^*)},
\end{equation}
\begin{equation}
\label{a2}
D^{T*}[2]=\frac{\nu_4^*(\chi-\mu-\chi^2(\nu_3^*/\nu_7^*))-\nu_2^*\chi(1-\chi(\nu_6^*/\nu_7^*))}
{\nu_1^*\nu_4^*-\nu_2^*(\nu_5^*-\zeta^*)},
\end{equation}
\begin{equation}
\label{a3}
D^{*}[2]=-\frac{\mu\nu_4^*}
{\nu_1^*\nu_4^*-\nu_2^*(\nu_5^*-\zeta^*)},
\end{equation}
where
\begin{equation}
\label{a4}
\nu_{1}^*=\frac{2\pi^{(d-1)/2}}{d\Gamma\left(\frac{d}{2}\right)}\left(\frac{\overline{\sigma}}{\sigma}
\right)^{d-1}{\cal M}(1+\alpha_0) \left(\frac{1+\theta}{\theta}\right)^{1/2},
\end{equation}
\begin{equation}
\label{a5} \nu_{2}^*=\frac{\pi^{(d-1)/2}}
{d\Gamma\left(\frac{d}{2}\right)}\left(\frac{\overline{\sigma}}{\sigma}\right)^{d-1}
{\cal M}(1+\alpha_0)[\theta(1+\theta)]^{-1/2},
\end{equation}
\begin{equation}
\label{a6} \nu_{3}^*=-\frac{\pi^{(d-1)/2}}
{d\Gamma\left(\frac{d}{2}\right)}\left(\frac{\overline{\sigma}}{\sigma}\right)^{d-1}
\frac{{\cal M}^2}{{\cal M}_{0}}(1+\alpha_0) \theta^{5/2}(1+\theta)^{-1/2},
\end{equation}
\begin{eqnarray}
\label{a7}
\nu_{4}^*&=&\frac{\pi^{(d-1)/2}}
{d(d+2)\Gamma\left(\frac{d}{2}\right)}\left(\frac{\overline{\sigma}}{\sigma}\right)^{d-1}
{\cal M}(1+\alpha_0)\left(\frac{\theta}{1+\theta}\right)^{3/2}\nonumber\\
&  & \times
\left[A-(d+2)\frac{1+\theta}{\theta} B\right],
\end{eqnarray}
\begin{equation}
\label{a8}
\nu_{5}^*=\frac{2\pi^{(d-1)/2}}
{d(d+2)\Gamma\left(\frac{d}{2}\right)}\left(\frac{\overline{\sigma}}{\sigma}\right)^{d-1}
{\cal M}(1+\alpha_0)\left(\frac{\theta}{1+\theta}\right)^{1/2}B,
\end{equation}
\begin{eqnarray}
\label{a9} \nu_{6}^*&=&-\frac{\pi^{(d-1)/2}}
{d(d+2)\Gamma\left(\frac{d}{2}\right)}\left(\frac{\overline{\sigma}}{\sigma}\right)^{d-1}
\frac{{\cal M}^2}{{\cal M}_{0}}(1+\alpha_0)\nonumber\\
& & \times \left(\frac{\theta}{1+\theta}\right)^{3/2}
\left[C+(d+2)(1+\theta)D\right],
\end{eqnarray}
\begin{eqnarray}
\label{a10} \nu_7^*&=&\frac{8}{d(d+2)}\frac{\pi^{(d-1)/2}}{\sqrt{2}\Gamma(d/2)}
(1+\alpha)\nonumber\\
& & \times\left(\frac{d-1}{2}+\frac{3}{16}(d+8)(1-\alpha)\right).
\end{eqnarray}
Here, ${\cal M}=m/(m+m_0)$, ${\cal M}_0=m_0/(m+m_0)$ and $\theta=m_0T/mT_0\equiv \mu/\chi$ is the mean-square velocity of the gas particles relative to that of the impurity particle. In addition, $\zeta^*$ is given by Eq.\ \eqref{3.5} and in Eqs.\ \eqref{a7}--\eqref{a9} we have introduced the quantities
\begin{widetext}
\begin{eqnarray}
\label{a11} A&=&2{\cal M}^2\left(\frac{1+\theta}{\theta}\right)^{2}
\left(2\alpha_0^{2}-\frac{d+3}{2}\alpha_0+d+1\right)
\left[d+5+(d+2)\theta\right] -{\cal M}(1+\theta) \left\{\lambda\theta^{-2}[(d+5)+(d+2)\theta]\right.
\nonumber\\
& & \left.\times [(11+d)\alpha_0
-5d-7]-\theta^{-1}[20+d(15-7\alpha_0)+d^2(1-\alpha_0)-28\alpha_0] -(d+2)^2(1-\alpha_0)\right\}
\nonumber\\
& & +3(d+3)\lambda^2\theta^{-2}[d+5+(d+2)\theta]+ 2\lambda\theta^{-1}[24+11d+d^2+(d+2)^2\theta]
+(d+2)\theta^{-1} [d+3+(d+8)\theta]
\nonumber\\
& &-(d+2)(1+\theta)\theta^{-2}
[d+3+(d+2)\theta],
\end{eqnarray}
\begin{eqnarray}
\label{a12} B&=& (d+2)(1+2\lambda)+{\cal M}(1+\theta)\left\{(d+2)(1-\alpha_0)
-[(11+d)\alpha_0-5d-7]\lambda\theta^{-1}\right\}+3(d+3)\lambda^2\theta^{-1}\nonumber\\
& & +2{\cal M}^2\left(2\alpha_0^{2}-\frac{d+3}{2}\alpha
_{12}+d+1\right)\theta^{-1}(1+\theta)^2- (d+2)\theta^{-1}(1+\theta),
\end{eqnarray}
\begin{eqnarray}
\label{a13} C&=&2{\cal M}^2(1+\theta)^2
\left(2\alpha_0^{2}-\frac{d+3}{2}\alpha_0+d+1\right)
\left[d+2+(d+5)\theta\right]-{\cal M}(1+\theta) \left\{\lambda[d+2+(d+5)\theta]\right.\nonumber\\
& & \left.\times [(11+d)\alpha_0-5d-7] +\theta[20+d(15-7\alpha_0)+d^2(1-\alpha_0)-28\alpha_0] +(d+2)^2(1-\alpha_0)\right\}
\nonumber\\
& & +3(d+3)\lambda^2[d+2+(d+5)\theta]- 2\lambda[(d+2)^2+(24+11d+d^2)\theta]+(d+2)\theta [d+8+(d+3)\theta]
\nonumber\\
& & -(d+2)(1+\theta)[d+2+(d+3)\theta], \nonumber\\
\end{eqnarray}
\begin{eqnarray}
\label{a14} D&=& (d+2)(2\lambda-\theta)+{\cal M}(1+\theta)\left\{(d+2)(1-\alpha_0)
+[(11+d)\alpha_0-5d-7]\lambda\right\}-3(d+3)\lambda^2\nonumber\\
& &-2{\cal M}^2\left(2\alpha_0^{2}-\frac{d+3}{2}\alpha _{12}+d+1\right)
(1+\theta)^2+(d+2)(1+\theta),
\end{eqnarray}
\end{widetext}
where $\lambda={\cal M}_{0}(1-\chi^{-1})$.

\bibliographystyle{apsrev}
\bibliography{SHG}

\end{document}